\documentclass[aps,prl,reprint,twocolumn,longbibliography,superscriptaddress,draft=false]{revtex4-2}
\usepackage{hyperref}
\hypersetup{colorlinks=true,
		    linkcolor=blue,
		    citecolor=blue,
		    allcolors=blue}
\usepackage{natbib}
\usepackage{amsmath, amsfonts, bm, bbm, braket, mathtools}
\usepackage{array}
\usepackage{graphicx,overpic}
\graphicspath{{./}}
\usepackage{pifont}% http://ctan.org/pkg/pifont
\usepackage{lineno}
\newcommand{\mybullet}{\Large$\circ$}
\newcommand{\mycross}{$\times$}

\begin{document}

%\preprint{}

\title{Determining Kitaev interaction in spin-$S$ honeycomb Mott insulators }
\author{Jiefu Cen}
\affiliation{Department of Physics, University of Toronto, Toronto, Ontario, Canada M5S 1A7}
\author{Hae-Young Kee}
\email[]{hykee@physics.utoronto.ca}
\affiliation{Department of Physics, University of Toronto, Toronto, Ontario, Canada M5S 1A7}
\affiliation{Canadian Institute for Advanced Research, CIFAR Program in Quantum Materials, Toronto, Ontario, Canada, M5G 1M1}
\begin{abstract}
The Kitaev interaction in a honeycomb lattice with higher-spin $S$ %> \frac{1}{2}$ 
has been one of the central attractions, as it may offer quantum spin liquids.
%However,  since the Hund\rq{}s coupling necessary for higher-spin and the spin-orbit coupling (SOC) required for the Kitaev interaction are not compatible to each other, %
%it was unclear how to generate the higher-spin Kitaev interaction until recently. 
A microscopic theory showed that when the Hund\rq{}s coupling at the transition metal generates $S> \frac{1}{2}$, the spin-orbit coupling at the heavy ligands provides a route to the Kitaev interaction. However, there have been debates over its strength compared to other symmetry-allowed interactions. Investigating the symmetry of the Hamiltonian for general $S$, we show the magnon energies at two momentum points related by a broken mirror symmetry reflect the Kitaev interaction when a magnetic field is in the mirror plane. Applying the symmetry analysis to CrI$_3$ with $S=\frac{3}{2}$ together with the available angle-dependent ferromagnetic resonance data, we estimate the Kitaev interaction out of the full Hamiltonian %including single-ion anisotropy,  Heisenberg interaction within and between the layers, and DM interaction. %
and find that it is sub-dominant. % We further predict that the magnon energies at two high-symmetric momentum points related by the mirror plane broken by the Kitaev interaction reflects its presence when the magnetic field is in the mirror plane.
Our theory can be tested by inelastic neutron scattering on candidate materials under the proposed magnetic field direction, which will advance the search for general $S$ Kitaev materials.
\end{abstract}
%\date{\today}
\maketitle

{\it Introduction} --
The Kitaev interaction in a honeycomb lattice has attracted much attention as it offers the exactly solvable Kitaev quantum spin liquid for spin $S=\frac{1}{2}$ \cite{kitaev2006,Jackeli2009PRL,Rau2014PRL,singh2012relevance,choi2012prl,plumb2014prb,modic2014realization,HSKim2015prb,sears2015prb,sandilands2015continuum,johnson2015monoclinic,banerjee2016proximate,HSKim2016structure}. The higher-spin Kitaev model has been theoretically studied \cite{baskaran2007dynamics,Baskaran2008SpinS,Oitmaa2018SpinS}, and numerical results have suggested that its ground state exhibits a quantum spin liquid \cite{Koga2018Spin1Ground,Zhu2020Spin1Ground,Hickey2020Spin1,Khait2021Spin1,Chen2022Spin1}. Recently, a microscopic theory to realize higher S in two-dimensional (2D) materials was proposed. The bond-dependent Kitaev and symmetric off-diagonal interactions denoted by $\Gamma$ and $\Gamma^\prime$ were derived using the strong coupling approach \cite{Peter2019HigherK}. These interactions originate from the strong spin-orbit coupling (SOC) at the heavy ligands together with the Hund's coupling at transition metal sites. 
Shortly after, it was suggested that CrI$_3$ with ferromagnetic (FM) ordering at low temperature is a Kitaev candidate material with $S=\frac{3}{2}$ where the Kitaev interaction is dominant over other symmetry-allowed interactions \cite{Lee2020PRL}. 
However, its strength compared to other interactions such as the most conventional Heisenberg term has been much debated \cite{Chen2018Neutron,Chen2020Neutron}. %This calls for a  to determine the Kitaev interaction.
%Another bond-dependent interaction called $\Gamma$ is responsible for the spin gap. 
%arising from the strong SOC provided by the heavy ligand iondine are important to 
%They are allowed by the symmetry and thus expected to be present in CrI$_3$.
 %which are allowed by the symmerty of the lattice \cite{todo}.   
 %were estimated by ab initio calculations using a strong-coupling perturbation theory \cite{todo}. 
 %The ferromagnetic Heisenberg interaction was found to be dominant, and the single-ion anisotropy contributed the most to the magnetic anisotropy.
 
Here we study the extended Kitaev model containing all the symmetry-allowed interactions with spin $S$ in a honeycomb lattice to address the significance of the Kitaev interaction. 
Based on the symmetry analysis in momentum space, we show that the magnon energies at two momenta related by a broken mirror symmetry reflect the presence of the bond-dependent interactions including the Kitaev, when a magnetic field is applied in the broken mirror plane. 
%
%a way to estimate its size by utilizing the magnetic field in special directions associated with the broken mirror symmetry that reflects the presence of the Kitaev interaction. %Assuming that $\Gamma$ interaction is tiny in higher-spin as shown in Ref. \cite{} and 
Applying the theory to CrI$_3$ together with the currently available angle-dependent ferromagnetic resonance (FMR) data,
%energies while sweeping certain field angles. This difference is due to the lack of $\pi$-rotation along a particular axis which is  %$C_{2a}$ symmetry in the model 
%broken by the Kitaev interaction. % when $\Gamma$ and $\Gamma^{\prime}$ are small \cite{todo}. 
%We find that the finite DM interaction with the out-of-plane d-vector does not change the symmetry analysis.% 
we find the Kitaev interaction to be sub-dominant, i.e., the second largest after the Heisenberg interaction.
%and approximately 40\% of the Heisenberg interaction.  
%We predict the magnon energies at two momenta related by the broken mirror symmetry associated with
Our proposal can be tested by inelastic neutron scattering (INS) under the magnetic field in the broken mirror plane. % in candidate honeycomb materials. 
The suggested experimental setup can be extended to general models for any $S$.
%as long as $\Gamma$ and $\Gamma'$ interactions are negligible, or $\Gamma \sim \Gamma'$. 

 %We will then apply the theory and estimate the size of Kitaev intearction in CrI$_3$ together with available angle-dependent ferromagnetic resonance (FMR) data and the proposal in Ref. \cite{Cen2022}. 
%For $S=1/2$, the significant $\Gamma$ interaction requires further consideration. Additional information such as the spin gap is required. 

\begin{figure} 
\includegraphics[width=1.0\linewidth,trim={0mm 00mm 0mm 00mm}]{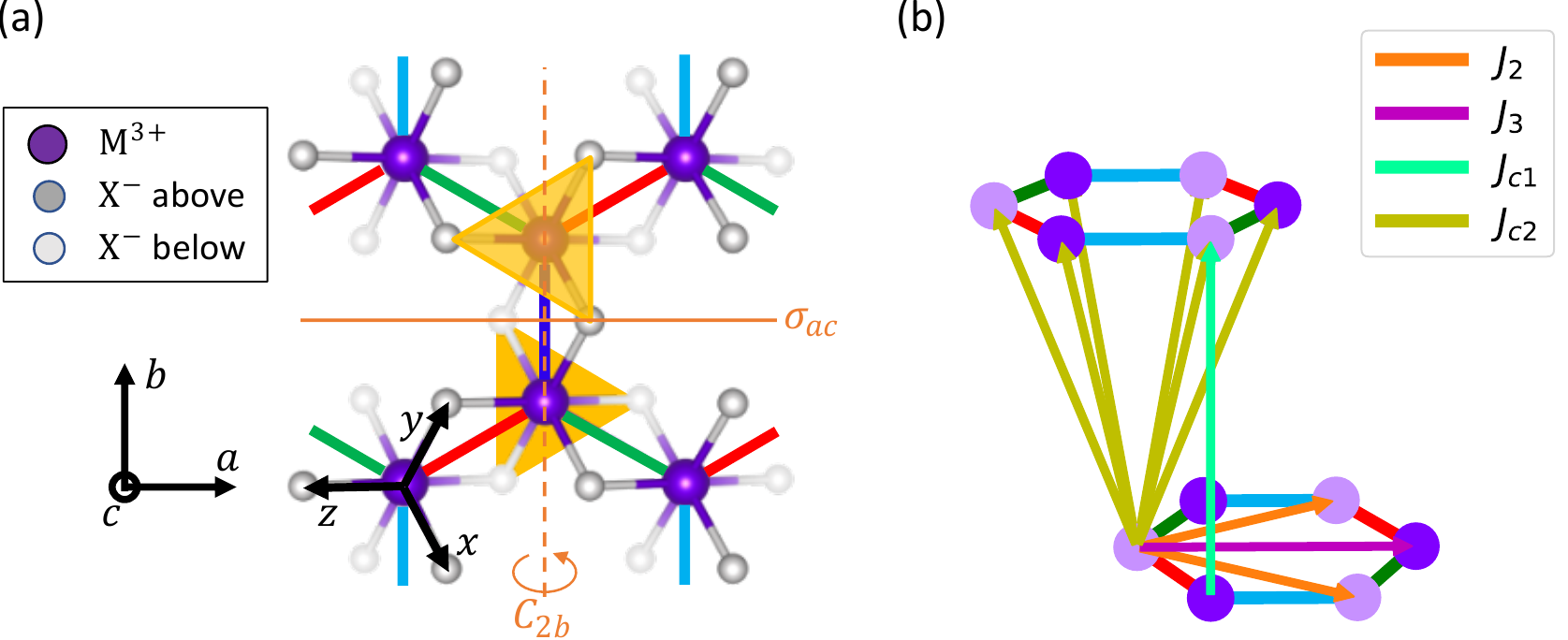}
\caption{
%\textbf{Crystal structure and further Heisenberg interactions.} 
(a) Schematic of the honeycomb lattice of transition metal ions (purple) in edge sharing octahedra environment of anions (above the honeycomb plane: gray,  below the plane: light gray).  Octahedral $\boldsymbol{x}\boldsymbol{y}\boldsymbol{z}$ axes, crystallographic $\boldsymbol{a}\boldsymbol{b}\boldsymbol{c}$ axes, and the Kitaev bonds x (red), y (green), z (blue) are indicated. 
%$C_{2a}$ and $C_{2b}$ symmetries (orange) are highlighted. The octahedra  environment breaks $C_{2a}$, while $C_{2b}$ symmetry is intact. 
(b) further neighbour $(J_2, J_3)$ and interlayer $(J_{c1}, J_{c2}$) Heisenberg interactions denoted by $J_n$ in Eq. (1) are included.}
\label{Fig1}
 \end{figure}

{\it Spin Model for general spin $S$} --
Let us begin with the general Hamiltonian and inspect its symmetry.
%
%We review the strategy based on symmetry in Ref. \cite{todo} that is used to obtain the Kitaev interaction in our model.  
The spin exchange Hamiltonian for spin $S$ on the ideal honeycomb lattice is known as the $JK\Gamma$ model \cite{Rau2014PRL}, where $J$, $K$ and $\Gamma$ are Heisenberg, Kitaev and symmetric off-diagonal interactions respectively. When trigonal distortion is introduced, the single-ion anisotropy and another off-diagonal interaction denoted by $\Gamma'$ are also present\cite{Rau2014-arxiv}. Transforming the octahedral $\boldsymbol{x}\boldsymbol{y}\boldsymbol{z}$ axes to the crystallographic %relevant for spin-$\frac{3}{2}$ CrI$_3$ in the $\boldsymbol{a}-\boldsymbol{b}-\boldsymbol{c}$ 
%
%\begin{equation}\begin{split}
%\mathcal{H} =& \sum_{\langle ij\rangle\in\alpha\beta(\gamma)} \Big[ J{\bf S}_{i}\cdot {\bf S}_{j}+KS_{i}^{\gamma}S_{j}^{\gamma}+\Gamma(S_{i}^{\alpha}S_{j}^{\beta}+S_{i}^{\beta}S_{j}^{\alpha}) \\
%&\qquad\qquad+\Gamma^{\prime}(S_{i}^{\alpha}S_{j}^{\gamma}+S_{i}^{\gamma}S_{j}^{\alpha}+S_{i}^{\beta}S_{j}^{\gamma}+S_{i}^{\gamma}S_{j}^{\beta}) \Big] \\
%&+\sum_{i}A_{c}(\vec{S}_{i}\cdot\hat{c})^{2}+\sum_{\langle\langle i,j\rangle\rangle}\vec{D}_{c}\cdot(\vec{S}_{i}\times\vec{S}_{j}).
%\end{split}
%\end{equation}
%To analyze the symmetries of the Hamiltonian, the $JK\Gamma\Gamma^\prime$ Hamiltonian in the octahedral axes is rewritten in the 
$\boldsymbol{a}\boldsymbol{b}\boldsymbol{c}$ 
axes shown in Fig. \ref{Fig1}(a), the general Hamiltonian beyond nearest neighbor (n.n.) is given by \cite{Onoda2011,Ross2011JcpJpp,Chaloupka2015hidden,Cen2022}
\begin{eqnarray}\label{eq:hamiltonian}
%\begin{split}
\mathcal{H}&=&\sum_{\langle i,j\rangle}\Bigg[J_{XY}(S_{i}^{a}S_{j}^{a}+S_{i}^{b}S_{j}^{b})+J_{Z}S_{i}^{c}S_{j}^{c}\nonumber\\
&&+ J_{ab}\left[c_{\phi_{\gamma}}(S_{i}^{a}S_{j}^{a}-S_{i}^{b}S_{j}^{b})-s_{\phi_{\gamma}}(S_{i}^{a}S_{j}^{b}+S_{i}^{b}S_{j}^{a})\right]\nonumber\\
&&-\sqrt{2}J_{ac}\left[c_{\phi_{\gamma}}(S_{i}^{a}S_{j}^{c}+S_{i}^{c}S_{j}^{a})+s_{\phi_{\gamma}}(S_{i}^{b}S_{j}^{c}+S_{i}^{c}S_{j}^{b})\right] \Bigg] %+ \mathcal{H}^\prime
\nonumber\\
&& +\sum_{i}A_{c}(S_{i}^{c})^{2}
+\sum_{\langle\langle i,j\rangle\rangle}D_{c}\text{ sgn}(ij)(S_{i}^{a}S_{j}^{b}-S_{i}^{b}S_{j}^{a})
\nonumber\\
&&+ \sum_{(i,j)_n} J_{\text{n}}{\bf S}_{i}\cdot {\bf S}_{j},
%\end{split}
\end{eqnarray}
where
$c_{\phi_{\gamma}} = \cos\phi_\gamma$, $s_{\phi_{\gamma}} = \sin\phi_\gamma$ and $\phi_\gamma = 0,\frac{2\pi}{3}$ and $\frac{4\pi}{3}$ for $\text{\ensuremath{\gamma=\text{z-, x- and y}}}$-bond respectively. 
$A_c$ and $D_c$ are single-ion anisotropy due to the trigonal distortion and the second n.n. Dzyalonshinskii-Moriya (DM) interaction with the D-vector along the c-axis \cite{Chen2020Neutron}, respectively. 
%$A_c$ is single-ion anisotropy due to the trigonal distortion, and $D_c$ is the second n.n. Dzyalonshinskii-Moriya (DM) interaction with the D-vector along the c-axis (sgn$(ij)$ is the sign convention for the cross product). 
$J_{\text{n}}$ are Heisenberg interactions between further neighbors and interlayers denoted by $(i,j)_n$, as shown in Fig. \ref{Fig1} (b). They are much smaller than the n.n. interactions.

The bond-dependent exchange interactions, $J_{ab}$ and $J_{ac}$ with $\phi_{\gamma}$ are related to %can be rewritten in x-y-z octahedron coordinate together with $J_{XY}$ and $J_Z$, and related to 
J, K, $\Gamma$ and %another symmetric off-diagonal interaction 
$\Gamma^\prime$ as follows: %introduced by the trigonal distortion as follows.
\begin{eqnarray}%\begin{split}
J_{XY}&=&J+ J_{ac} - \Gamma^\prime, \;\;%\frac{1}{3}K -\frac{1}{3}\Gamma-\frac{2}{3}\Gamma^{\prime},\;\; 
J_{Z} =J+J_{ab} + 2 \Gamma^\prime,% \frac{1}{3}K+\frac{2}{3}\Gamma+\frac{4}{3}\Gamma^{\prime},
\nonumber\\
J_{ab}&=&\frac{1}{3}K+\frac{2}{3}(\Gamma-\Gamma^{\prime}), \;\; J_{ac}=\frac{1}{3}K-\frac{1}{3}(\Gamma-\Gamma^{\prime}).
%\end{split}
\end{eqnarray}
A combination of $\Gamma$ and $\Gamma^{\prime}$ leads to the in- and out-of-plane magnetic anisotropy and the spin gap at zero momentum since $J_Z - J_{XY} = \Gamma + 2 \Gamma'$. The single-ion anisotropy $A_c$ also generates the anisotropy and spin gap except $S=\frac{1}{2}$, for which it is a constant.
Below we will first analyze the symmetry of the Hamiltonian. %similar to the earlier work for $S=\frac{1}{2}$ \cite{Cen2022}. The advantage of the current proposal compared to the one in Ref. \cite{Cen2022} will be discussed later.

{\it Symmetry Analysis} -- 
The symmetries of the Hamiltonian were analyzed for $S=\frac{1}{2}$ in Ref. \cite{Cen2022}. For general $S$ including combined symmetry operations, they are summarized in the first row of Table \ref{table1}. 
The cross and circle refer to broken and preserved symmetries, respectively.
$\mathcal{T}$ is the time reversal symmetry $\mathcal{T}: \bold{S}\rightarrow -\bold{S}$. $C_{3c}$ is the $\frac{2\pi}{3}$ rotation of the lattice around the $\hat{c}$ axis. $C_{2a}$ and $C_{2b}$ are $\pi$ rotations about the $\hat{a}$ and $\hat{b}$ axes respectively.
The spatial inversion and translation symmetries are intact. Since the octahedron forms two triangles above and below the transition metal $M$ plane as shown in Fig. 1 (a), the mirror plane exists only in the $ac$ plane equivalent to $C_{2b}$ because of spatial inversion. The mirror symmetry about the $bc$ plane (i.e., $C_{2a}$) is broken, which results in a finite $J_{ac}$ in the Hamiltonian. The red crosses are the broken symmetries due to a finite $J_{ac}$, and the blue circles are the preserved symmetries if $J_{ac}$ was zero. %These symmetries can be probed under two magnetic field directions in the $ac$ plane to determine $J_{ac}$ as discussed in our earlier study \cite{Cen2022}.

 The second and third rows show how the symmetries are affected under special magnetic field directions. % are listed in Table \ref{table1}. 
When the magnetic field is applied, $\mathcal{T}$ is broken. $C_{3c}$ is also broken as the field is away from the c-axis. The combined operation $\mathcal{T}C_{2b}$ is preserved when the field is in the $ac$ plane, while it is broken when the field is in the $bc$ plane.
It is important to note that 
 another combined operation $\mathcal{T}C_{2a}$ is preserved if $J_{ac}$ was absent, as indicated by the blue circle. 
 %at the bottom corner of the table. 
%This difference would signal the finite $J_{ac}$.
Since the lack of $\mathcal{T}C_{2a}$ is only related to a finite $J_{ac}$ in $\mathcal{H}$ while all the other symmetries are broken by the field, 
the difference in the spin excitations between two momenta related by $\mathcal{T}C_{2a}$ is only due to $J_{ac}$, i.e., the magnon energy difference between these two momenta can be used to determine $J_{ac}$.
%
%Since the inversion symmetry is preserved, they correspond to the mirror planes of $bc$ and $ac$ respectively. 

\begin{table}
\centering
\begin{tabular}{ |c| >{\centering\arraybackslash} m{1cm} >{\centering\arraybackslash} m{1cm} >{\centering\arraybackslash} m{1cm} >{\centering\arraybackslash} m{1cm} >{\centering\arraybackslash} m{1cm}| } 
\hline
  & $C_{3c}$ & $C_{2b}$ & $\mathcal{T}C_{2b}$ & $C_{2a}$ & $\mathcal{T}C_{2a}$ \\ 
 \hline
 $\bold{h}=0$ & \mybullet &  \mybullet & \mybullet & {\color{red} \mycross}/{\color{blue} \mybullet} & {\color{red} \mycross}/{\color{blue} \mybullet} \\ 

 $\bold{h}$ in $ac$ plane & \mycross &  \mycross & \mybullet & \mycross & \mycross \\ 

 $\bold{h}$ in $bc$ plane & \mycross &  \mycross & \mycross & \mycross & {\color{red} \mycross}/{\color{blue} \mybullet} \\ 
%
%\hline
%\multicolumn{6}{c}{\vspace{1cm}}\\
%\hline
% $J_{ac}=0$ & $\mathcal{T}$ & $C_{2b}$ & $\mathcal{T}C_{2b}$ & $C_{2a}$ & $\mathcal{T}C_{2a}$ \\ 
 %\hline
 %$\bold{h}=0$ & \mybullet &  \mybullet & \mybullet & {\color{red} \mybullet} & {\color{red} \mybullet} \\ %
 %$\bold{h}$ in $a-c$ plane & \mycross &  \mycross & \mybullet & \mycross & \mycross \\ 
%
 %$\bold{h}$ in $b-c$ plane & \mycross &  \mycross & \mycross & \mycross & {\color{red} \mybullet} \\ 
%
\hline
\end{tabular}
\caption{Symmetries of the Hamiltonian under the magnetic field in specific planes. The cross and circle refer to broken and preserved symmetries respectively. The red cross indicates the broken symmetry due to a finite $J_{ac}$, and the blue circle the preserved symmetry if $J_{ac}$ was absent. We assume that the spatial inversion and translation symmetries are intact, and there is no spontaneous symmetry breaking due to magnetic ordering.}
\label{table1}
\end{table}

%We will present the symmetry of the general $S$ model and how to use the result for higher $S$ systems, and then discuss later how to estimate the Kitaev interaction for $S=1/2$ where $\Gamma$ interaction is large. 
%, and also refer to our earlier study to differentiate the contributions from $\Gamma$ and Kitaev interactions.\cite{} %symmetry analysis below. 
%For spin $S=1/2$ materials with significant $\Gamma$ interaction, the difference reflects the combination $J_{ac}=K/3 - (\Gamma-\Gamma^\prime)/3$, so we use the strategy as described in Ref. \cite{Cen2022} to estimate the Kitaev interaction. 

{\it A Simple Example} --
%{\it Inelastic neutron scattering experiments under magnetic fields} --
Based on the symmetry relation summarized in table I, 
we investigate the spin wave under a magnetic field in the $bc$ plane.
We expect that the difference in magnon energy between two momenta related by  $\mathcal{T}C_{2a}$ will signal the presence of $J_{ac}$. % \sim K$ when $\Gamma$ and $\Gamma'$ are ignored.

%further confirmations of our model, we also suggest 
%Similar to the FMR experiment, 

\begin{figure} [h]
\includegraphics[width=1.0\linewidth,trim={0mm 00mm 0mm 00mm}]{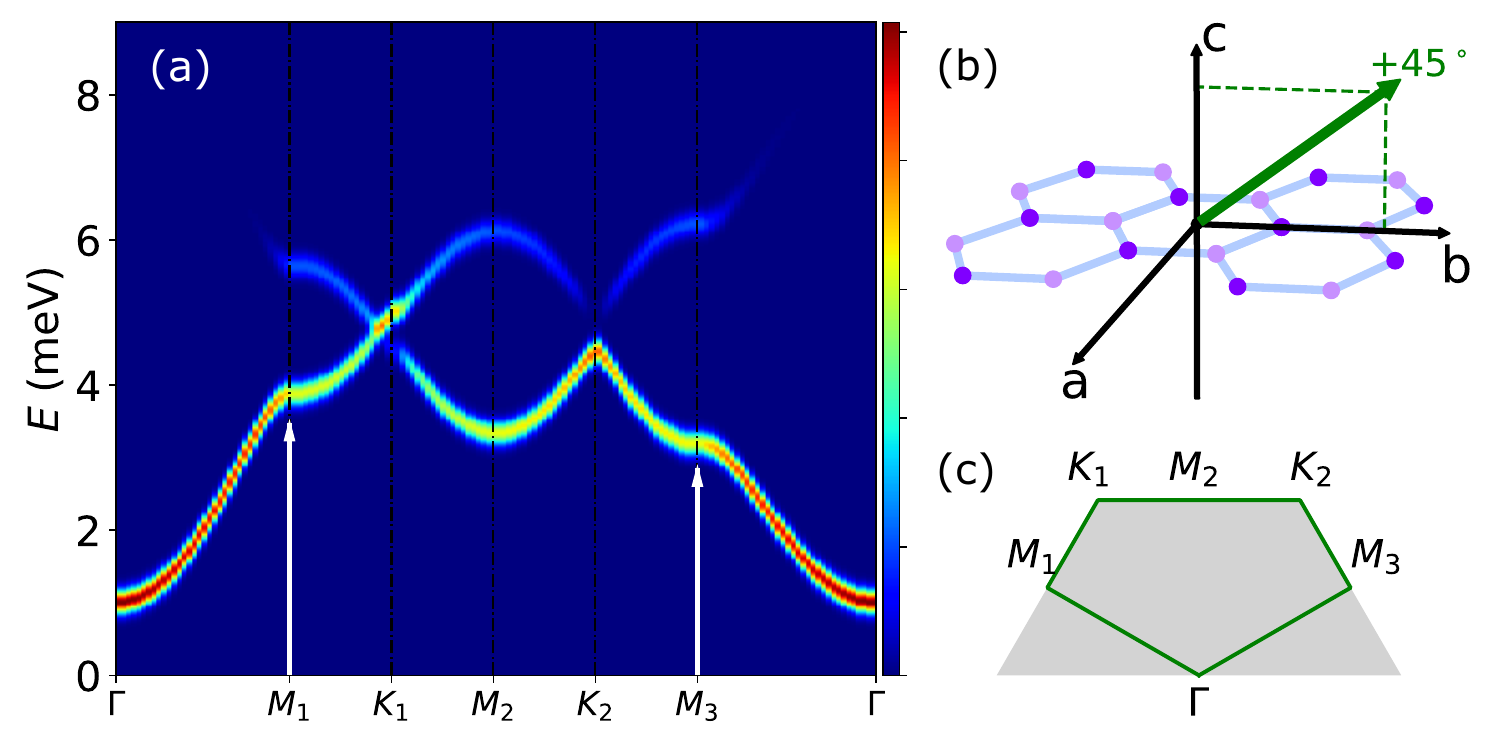}
\caption{Magnon spectrum with $J=-1$ and $K=0.5$ under a magnetic field in the $b-c$ plane with an angle of $\theta=45^\circ$ and the strength of $g\mu_B h=1$. The difference between $M_1$ and $M_3$ highlighted by the white arrows is due to the Kitaev interaction.}
\label{Fig2}
 \end{figure}

%An alternative way to detect the difference due to the Kitaev interaction is to put the magnetic field in the $b-c$ plane as argued in the symmetry analysis above. 
As an example, we check the simple case of $J=-1$, $K=0.5$, and set $g\mu_B h = 1$. 
The spin wave %calculation for our model with field strength $B=8.6$T(?) and 
with the field in the $bc$ plane with an angle of $\theta=45^\circ$ is shown in Fig. \ref{Fig2}(a)-(b). 
The momentum paths $\Gamma-M_1-K_1$ and $\Gamma-M_3-K_2$ shown Fig. 2(c) are related by the combination $\mathcal{T}C_{2a}$.  The excitations along these two paths are different since $C_{2a}$ reflects the nonzero Kitaev interaction due to the broken $bc$-plane mirror symmetry. 
For example, the white arrows indicate the significant difference in the energies between $M_1$ and $M_3$ (see the Supplemental Material for detailed analysis). However, the paths $K_1-M_2$ and $K_2-M_2$ have the same energies because the spatial inversion symmetry ($\bold{k}\rightarrow -\bold{k}$) is still intact. 
%Future inelastic neutron scattering (INS) experiments under magnetic fields that can verify the importance of the Kitaev interaction. 

While the current analysis of measuring the finite $J_{ac}$ works for any spin $S$, estimating the Kitaev interaction out of $J_{ac}$, which is a combination of $K$, $\Gamma$ and $\Gamma'$, depends on the size of $S$.
 For $S > 1/2$, it was shown that $\Gamma=0$ up to the 4th order perturbation in d$^8$ systems like Ni$_2$ with S=1\cite{Peter2019HigherK} , and $\Gamma,\Gamma^\prime \ll A_c$ and $\Gamma \sim \Gamma'$ 
 resulting in $J_{ac} \sim K/3$ for CrI$_3$ with $S=\frac{3}{2}$ \cite{StavropoulosPRR2021}. %This is rather expected, as $\Gamma (S^+_i S^+_j - S^-_i S^-_j)$ requires flipping two spins to get back to its original state, which occurs only in higher order perturbations beyond the second order in $S > 1/2$. $\Gamma^\prime$ is even smaller as it further requires the trigonal distortion. Thus we will use $A_c$ to account for the anisotropy and set  $\Gamma$ and $\Gamma^\prime$ roughly zero for higher-$S$ models. 
 On the other hand, for $S = \frac{1}{2}$, $\Gamma$ is significant while $\Gamma'$ can be ignored when trigonal distortion is minimal \cite{Rau2014-arxiv}. %Since there have been debates over the size of Kitaev and $\Gamma$, while, and 
Thus one has to subtract the $\Gamma$ contribution from $J_{ac}$ to determine the Kitaev interaction. Since the in- and out-of-plane anisotropy mainly originates from $\Gamma$ when there is no $A_c$ for $S=\frac{1}{2}$, one can first determine $\Gamma$ from the anisotropy and subtract $\Gamma$ from $J_{ac}$ to finalize the Kitaev. 
For $\alpha$-RuCl$_3$ \cite{HSKim2016structure,winter2016challenges,winter2017review,Janssen2019review}, a large effect of $J_{ac}$ is expected in the magnon energy difference between $M_1$ and $M_3$, as the Kitaev and $\Gamma$ have opposite signs making $J_{ac}$ large.

{\it Application to CrI$_3$} --
%
%Two-dimensional (2D)  magnetism in 
%atomically thin 2D van der Waals (vdW) materials has attracted a lot of interest as they offer a playground to explore exotic phases such as quantum spin liquids\cite{kitaev2006,cjk2010prl,banerjee2016proximate,baek2017evidence,Banerjee2018excitations,Takagi2019review} and their potential applications in spintronics \cite{AlexanderPRC2018,JiwuAIP2018,Burch2DvdWReview,Gibertini2DmagReview}. 
%Among them,  t chromium trihalides 
CrI$_3$ is a well-known quasi-2D van der Waals ferromagnet with the spin-$\frac{3}{2}$ Cr$^{3+}$ ions sitting on honeycombs of edge sharing octahedra. 
%Bulk CrI$_{3}$ has a critical temperature $T_{C}\sim61K$ \cite{todo}, and 
The FM order in bulk CrI$_3$ persists to single-layer with a reduced transition temperature $T_{C}\sim45$K \cite{Huang2017MonolayerCrI3}, and there has been a lot of attention on the minimal 2D spin model.
%Since a finite spin gap is required to stabilize 2D ordering at finite temperature, the origin of the magnetic anisotropy has been investigated.
%For example, the XXZ model with single-ion anisotropy was proposed.\cite{} Another possibility is the bond-dependent interaction such as Kitaev $(K)$ and $\Gamma$ interactions arising from the strong spin-orbit coupling provided by the heavy ligand iodine.  They are allowed by the symmetry and
%and thus expected to be present in CrI$_3$.   have been estimated by ab initio calculations using a strong-coupling perturbation theory \cite{todo}.   Dominant ferromagnetic Heisenberg interaction and sub-dominant Kitaev interaction were found, and the single-ion anisotropy contributed the most to the magnetic anisotropy.
%
In particular, the strength of the Kitaev interaction in CrI$_3$ has been debated. %on the spin model for CrI$_3$, especially the importance of the Kitaev interaction. 
Based on the angle-dependent FMR data \cite{Lee2020PRL},  Lee et al proposed that the Kitaev interaction is 25 times larger than the Heisenberg interaction ($|K|\sim25|J|$), and a tiny $\Gamma$ interaction is necessary to have a finite spin gap at the ordering wavevector. 
% a $K-J-\Gamma$ model. 
The FMR data were fitted with calculations using the Smit-Beljers-Suhl equation with the free energy obtained from the partition function of a mean-field Hamiltonian. %The proposed model has a dominant Kitaev interaction $|K|\sim25|J|$ \cite{todo}.
%Because mean-field method cannot account for the bond-dependent interactions $J_{ac}$ and $J_{ab}$ properly, the fit to the FMR data only estimated $J_X=J_Z=J+K/3$. 
$J$ and $K$ were resolved by the additional fit to the $T_c=61$K of the FM phase obtained by linear spin wave theory (LSWT). 
The large Kitaev interaction was also claimed to be responsible for the spin gap at the Dirac point. 
%In addition, they fit the gap at the Dirac $K$-point in a honeycomb lattice using the LSWT and claimed that the large Kitaev interactin was responsible for the spin gap at the Dirac point. 
 
On the other hand, Chen et al studied the magnon spectrum via the INS experiment under an in-plane magnetic field that aligns the spin moments in-plane \cite{Chen2020Neutron}.  In the FM ordered state, a well-defined magnon appeared. 
The earlier proposal with the dominant Kitaev interaction was ruled out because it did not fit the magnon spectrum under the in-plane field.
 Instead, they found that the FM Heisenberg interaction is dominant, and the significant DM interaction ($D_c$) with an out-of-plane D-vector defined on the second n.n. bond opens up the spin gap at the Dirac point.
 %($K_{1,2}$-point). 
 
 However, %the Kitaev interaction is not necessary to explain the INS data since %
 the measured INS data alone does not strongly constrain the strength of $K$, %and $|K|\le |J|$ is allowed,
because the contribution of $K$ to the spin gap is small compared to $D_c$ due to the sizable difference in the pre-factors in LSWT. This calls for further investigation to determine $K$. 
%
%without a magnetic field supports a similar $K-J-\Gamma$ model or a $J-A_c-D_c$ model based on linear spin wave theory (LSWT).  However, the $K-J-\Gamma$ model was ruled out by the spin-wave spectra with an in-plane magnetic field that aligns the moment in-plane. In the $J-A_c-D_c$ model, $J$ is dominant and $D_c$, the Dzyaloshinkii Moriya (DM) interaction with an out-of-plane vector, opens up the spin gap at the Dirac point. 
%While the Dirac gap can be generated by both the Kitaev and DM interactions, the recent LSWT theory together with the INS under in-plane field.
As discussed above, for $S > \frac{1}{2}$, $\Gamma$ and $\Gamma'$ are negligible, and $J_{ac} \sim K/3$. Since it is the Kitaev interaction that reflects the broken $bc$ mirror symmetry, we can apply the symmetry analysis to estimate the Kitaev interaction in CrI$_3$. However, there is no experiment on CrI$_3$ under the $bc$-plane field, so we will first determine the spin interactions using the available angle-dependent FMR and INS experiments. Then, we will present the magnon spectrum reflecting the Kitaev interaction under the $bc$-plane field, which can be tested in future experiments.

%The angle-dependent FMR shown in Fig. 3 by up- and down-triangles are adopted from Ref. \cite{}.
%We use the 12-site cluster shown in the inset to compute the resonance energy using the exact diagonalization (ED) under the $ac$-plane field. The magnetic field is swept from the angle $\theta$ measured from the a-axis. 
%For $\theta >0$, i.e, from ${\hat a}$ to $$+{\hat c}$, the computed resonance energy is shown in the blue circle, while
%for $\theta <0$, i.e, from ${\hat a}$ to $-{\hat c}$-axis, it is denoted by the blue circles. The difference in the resonance energy at all $\theta$ except 0 and $90^\circ$ reflects the strength of $J_{ac} \sim 1/3 K$.\cite{} 
%
%
\begin{figure} 
\includegraphics[width=1.0\linewidth, trim={0mm 00mm 0mm 0mm}]{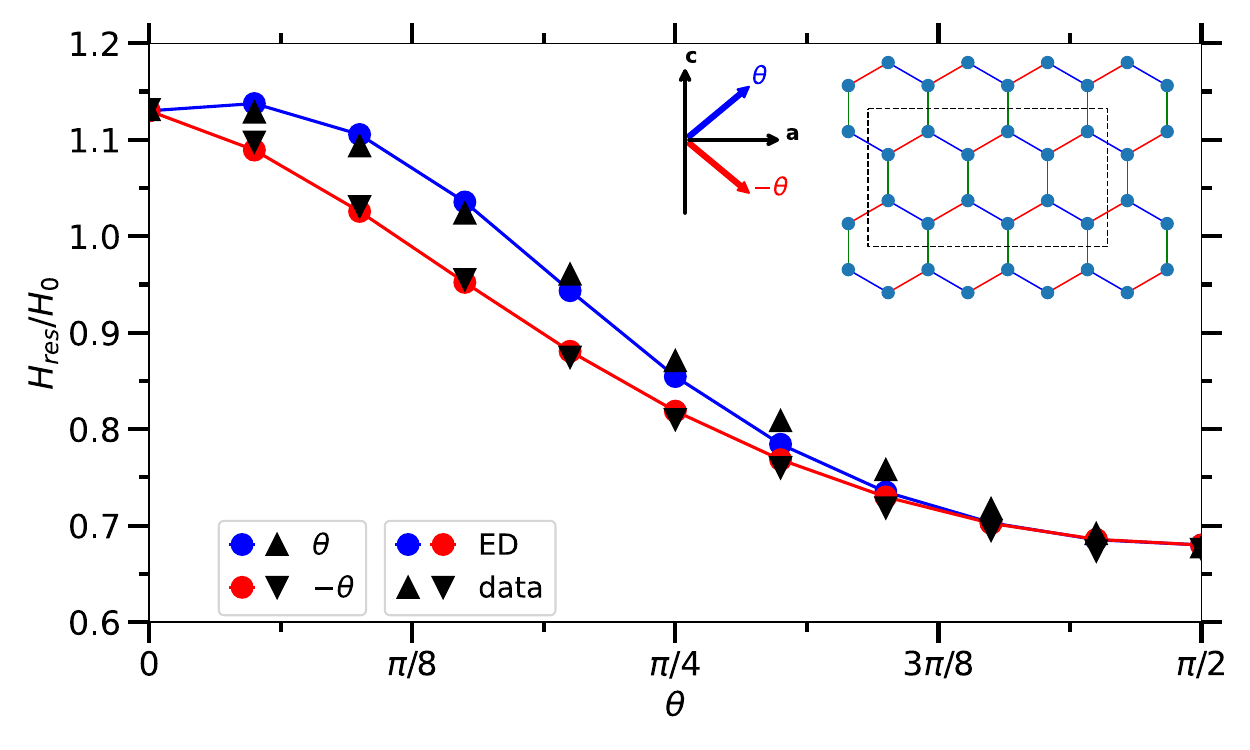}
\caption{FMR resonant field $H_{\text{res}}$ at 240 GHz resonant frequency as the field angle $\theta$ is varied, reproduce from Ref. \cite{Lee2020PRL}. $H_{\text{res}}$ is normalized by that of a free spin. The magnetic field angles are $\pm\theta$ in the $ac$ plane. Inset is the 12-site cluster used in the exact diagonalization (ED) of spin-$\frac{3}{2}$ to calculate the resonant field of our model. The calculation agrees with the FMR data well.}
\label{Fig3}
 \end{figure}
 
\begin{figure} [h]
\includegraphics[width=1.0\linewidth,trim={0mm 00mm 0mm 00mm}]{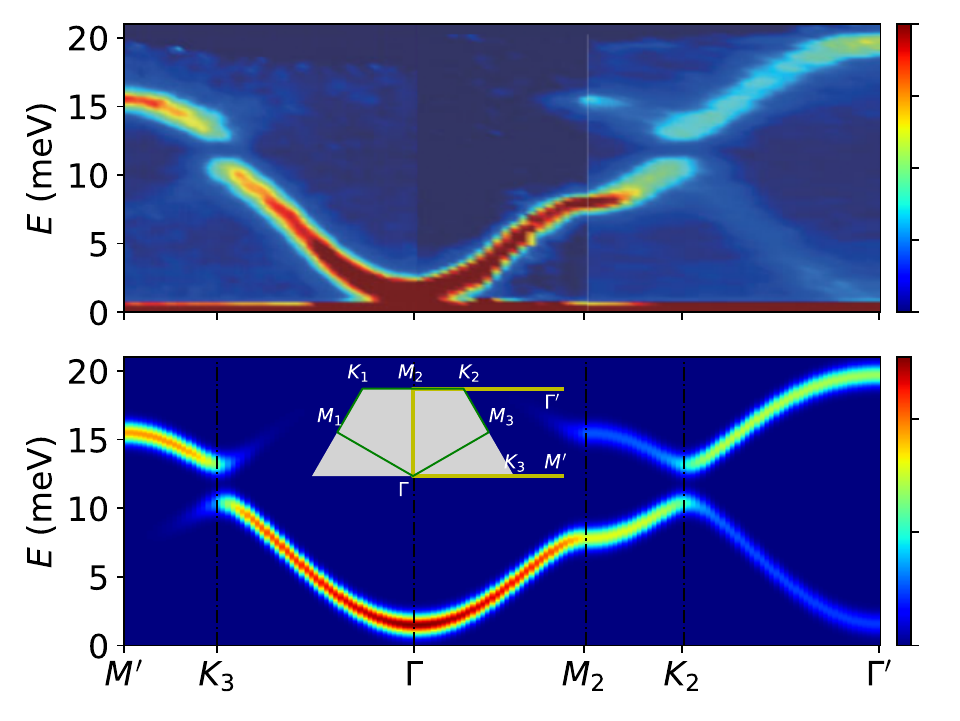}
\caption{(a) Inelastic neutron scattering (INS) data reproduced from Ref. \cite{Chen2020Neutron} and (b) Magnon spectrum of our model without a magnetic field using linear spin wave theory (LSWT). The momentum path (yellow) is shown in the inset. The calculation agrees with the INS data well.}
\label{Fig4}
 \end{figure}
 
We first obtain the relative strength of the Kitaev interaction  from the resonant magnetic field data in the FMR experiment %with 240 GHz resonant frequency at 5 K 
by Lee et al. \cite{Lee2020PRL}, where the magnetic field is applied in the $ac$ plane as shown in the inset of Fig. \ref{Fig3}. Thus we use the symmetry of the $ac$ plane proposed in Ref. \cite{Cen2022}. The field direction is in the $ac$ plane with angles $\theta$ ($0<\theta<\pi/2$) above and below the honeycomb plane. The data for $\theta$ and $-\theta$ are shown by up- and down-triangles respectively in Fig. \ref{Fig3}. Previous analysis of the FMR data using the free energy of a mean-field Hamiltonian could not account for bond-dependent interactions properly for spin excitations at zero momentum, so we use exact diagonalization (ED) \cite{Weie2008}. The resonant field at each field angle is calculated by matching the 240 GHz resonant frequency to the zero-momentum spin excitation in the ED of a 12-site cluster shown in the inset of Fig. \ref{Fig3}. 
The zero-momentum spin excitations have negligible finite-size effects with this magnetic field strength (see the Supplemental Material). In order to account for the difference in the resonant fields between $\theta$  and $-\theta$ as presented in Ref. \cite{Cen2022}, the relative strength between the Kitaev and Heisenberg interactions is required to be $|K|\sim0.4|J|$. The single-ion anisotropy $A_c$ and g-factor anisotropy $g_\perp=2.24$, $g_\parallel=1.69$ are determined from the in- and out-of-plane magnetic anisotropy of the resonant field. The calculated resonant field for our model is shown in Fig. \ref{Fig2} by blue ($\theta$) and red ($-\theta$) filled circles.

The Heisenberg $J$ and Kitaev $K$ interactions are then estimated using the INS data with no magnetic field by Chen et al. \cite{Chen2020Neutron}. We use LSWT since the spin excitations are well defined due to the dominant FM Heisenberg $J$. %energy is dominated by $J$ and $K$. 
The same small DM interaction $(D_c)$ with an out-of-plane D-vector and interlayer Heisenberg couplings $(J_{c1},J_{c2})$ as suggested by Chen et al. \cite{Chen2020Neutron} 
are used to account for the spin gap at the Dirac point and the out-of-plane momentum dependence respectively. Small second and third n.n. Heisenberg interactions $(J_2,J_3)$ are added to improve the fit, but they do not alter the main findings.
The INS data adopted from Ref. \cite{Chen2020Neutron} is shown in Fig. \ref{Fig4}(a) to make a comparison to the calculated spin wave spectrum without a field in Fig. \ref{Fig4}(b). The interlayer couplings are taken into account by integrating over the out-of-plane momentum. All the small Heisenberg interactions have negligible effects on the FMR resonant field calculation shown in Fig. \ref{Fig3}.  %Therefore, the importance of the Kitaev interaction can only be determined by the symmetry analysis for the FMR data, i.e. by comparing the spin excitations under two magnetic field directions with angles above and below the honeycomb plane. 

\begin{table}[h!]
\centering
\begin{tabular}{ |c|c|c|c| } 
 \hline
Interaction & value (meV) & Interaction & value (meV) \\ 
 \hline
 $J$ & -2.5 &  $J_2$ & -0.09 \\ 
 $K$ & 1.1 & $J_3$ & 0.13 \\ 
 $\Gamma$ & $\sim$0 &  $J_{c1}$ & 0.048  \\ 
 $\Gamma^\prime$ & $\sim$0 &  $J_{c2}$ & -0.071  \\ 
 $A_c$ & -0.23 & $D_c$ & 0.17 \\ 
 \hline
\end{tabular}
\caption{Spin exchange interactions that can accommodate both the FMR and INS experiments for CrI$_3$.}
\label{table2}
\end{table}

Table \ref{table2} lists the spin interactions obtained from the above analysis of the FMR and INS data.
Note the samples used in the INS and FMR experiments are different, but we expect them to have similar intralayer couplings since the Cr-Cr distances are similar. The interlayer couplings can vary a lot due to the different $c$-axis layer spacing, but they do not affect the broken $C_{2a}$ symmetry. Thus our method to determine the Kitaev interaction applies to both samples.
%, since the $C_{2a}$ symmetry. 
Here we assume the samples to be perfect single crystals. However, the mosaicity of the sample in the FMR experiment is unknown, and the effect of sample mosaicity will be discussed later. 

\begin{figure} [h]
\includegraphics[width=1.0\linewidth,trim={0mm 00mm 0mm 00mm}]{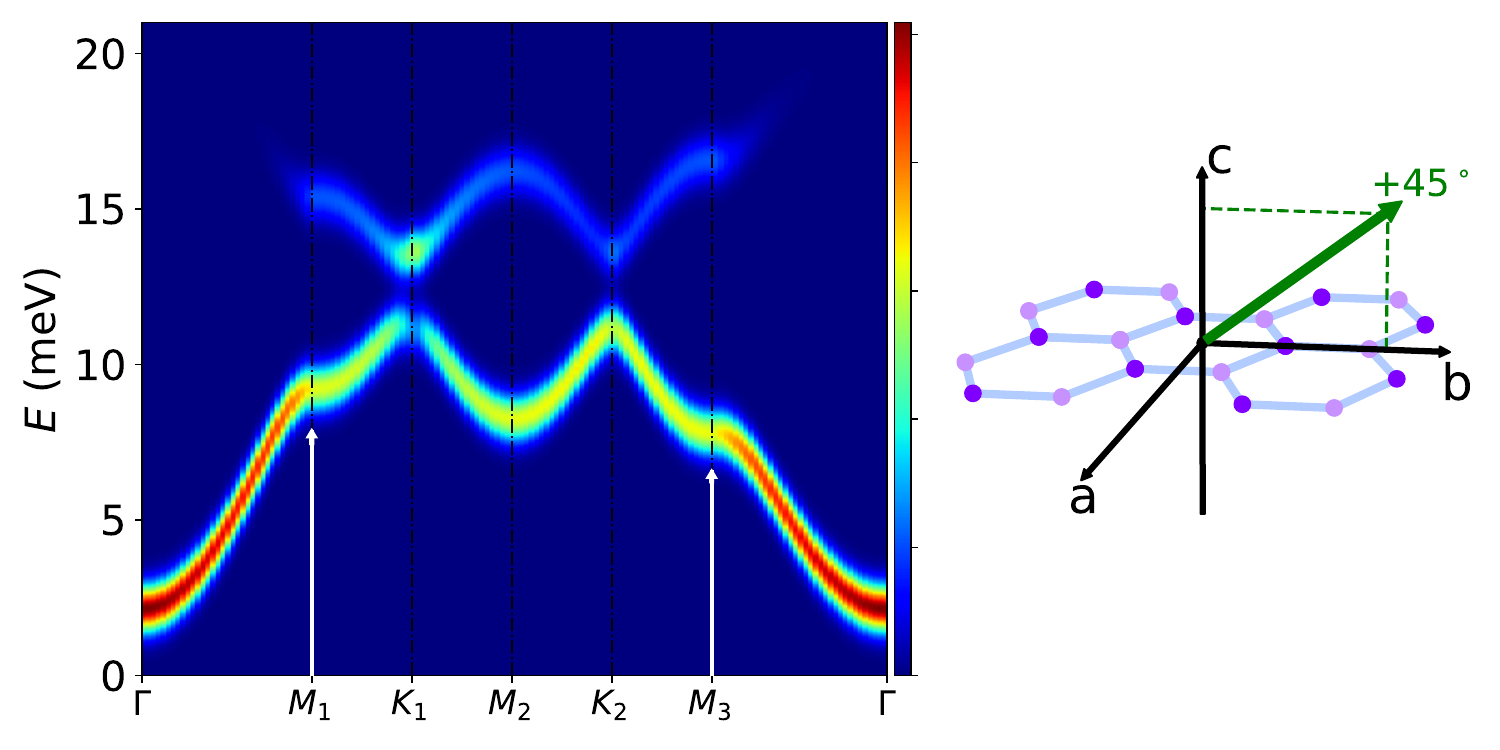}
\caption{Magnon spectrum of our model under a magnetic field in the $bc$ plane with an angle of $\theta=45^\circ$ and a field strength of 8.6T. The difference between $M_1$ and $M_3$ highlighted by the white arrows is due to the Kitaev interaction.}
\label{Fig5}
 \end{figure}

Using the obtained parameters of the spin model, we predict that the magnon spectrum under a $bc$-plane magnetic field should reflect the size of the Kitaev interaction. The result with a field angle of $\theta=45^\circ$ and a field strength of 8.6T is shown in Fig. \ref{Fig5}.
As expected, there is a noticeable difference in the magnon energies between $M_1$ and $M_3$. In the Supplemental Material, we also show the magnon spectrum under two $ac$-plane fields for the obtained CrI$_3$ model.

%\begin{figure} [h]
%\includegraphics[width=1.0\linewidth,trim={0mm 00mm 0mm 00mm}]{Fig5_INS_ac.pdf}
%\caption{Dynamic spin structure factor (DSSF) of the spin wave spectrum of our model under magnetic fields in the $a-c$ planes with angles $\theta=45^\circ$ (blue) and $\theta=-45^\circ$ (red). The differences between the two angles are due to the Kitaev interaction.}
%\label{Fig5}
% \end{figure}

%The spin waves can be measured under two magnetic field direction in the $a-c$ plane with angles $\theta$ above and below the honeycomb plane. Figure \ref{Fig4} emphasizes the differences in the spin waves with field strength $B=8.6$T and $\theta=45^\circ$ at all momenta due to the Kitaev interaction. Note the zero difference at the $\Gamma$ point is a defect of LSWT, as ED calculation has a nonzero difference. 

{\it Summary and Discussion} --
In summary, we show how to determine the Kitaev interaction for a general-$S$ model using the magnon energies at two momenta only related by the broken $bc$-plane mirror symmetry under the $bc$-plane field. We apply this method to CrI$_3$ with $S=\frac{3}{2}$ and show that it is a sub-Kitaev material, i.e, the Kitaev interaction is the second largest after the FM Heisenberg interaction. We predict its magnon spectrum that reflects the strength of the Kitaev interaction. Our method here requires INS measurement for a fixed magnetic field, which is advantageous over the method described in Ref. \cite{Cen2022} that needs INS measurements under rotated $ac$-plane fields, though it
%the method in \cite{Cen2022} also 
applies to zero-momentum excitations detected by
optical spectroscopies such as angle-dependant FMR. 

%The two proposed set-ups are complementary to each other when we consider sample mosaicity,
%which may affect the measured differences in energies related by the broken symmetry under magnetic fields. 
%Since $c$-axis mosaicity does not change the symmetries of the model, the difference between two momenta related by the broken $bc$-plane mirror symmetry still reflects the presence of $J_{ac}$. However, in-plane mosaicity leads to an extrinsic effect, because the field $bc$ plane may contain $a$-axis component, which contaminates the energy difference between the two momenta.
%On the other hand, using the two magnetic fields above and below the plane proposed in \cite{Cen2022}, in-plane mosaicity does not change the symmetry because the two fields are always related by $\mathcal{T}C_{2a}C_{2b}$, which reduces to $C_{2a}$ or $C_{2b}$ for fields in the $ac$- or $bc$-plane. However, $c$-axis mosaicity leads to an additional contribution to the difference between the two fields from the in- and out-of-plane magnetic anisotropy. %Therefore, two different proposals are needed to estimate the intrinsic effects and estimate the proper $J_ac$.

The relative strength of the Kitaev interaction for CrI$_3$ determined from the FMR data does not consider sample mosaicity. The effect of in-plane mosaicity can be estimated by applying a Gaussian filter in the in-plane field angle. The differences in the FMR resonances with the field angles above and below the honeycomb plane, which manifest the Kitaev interaction, are maximal when the field is in the $ac$ plane.
%and decrease to zero regardless of $K$ when the field is in the %($C_{3c}$-related) $bc$ plane.
Thus, Gaussian averaging the in-plane field angle will always decrease the resonance energy differences. The strength of the Kitaev interaction in CrI$_3$ could be larger if the FMR sample had in-plane mosaicity, which needs to be clarified in the future.

%Since there is no single-ion anisotropy for $S=1/2$, only $\Gamma$ contributes to the conventional in- and out-of-plane magnetic anisotropy in the case where $\Gamma^\prime$ is also negligible due to small trigonal distortion. Thus, $\Gamma$ can be estimated from the conventional anisotropy and subtracted from $J_{ac}$ to estimate $K$. If $\Gamma^\prime$ is significant, one needs to use the difference in energy at the Dirac point between field angles $\theta$ above and below the honeycomb plane in the $ac$-plane. Details are in Ref. \cite{Cen2022}. 

%\section*{Acknowledgements}
%
This work was supported by the Natural Sciences and Engineering Research Council of Canada and the Canada Research Chairs Program. HYK thanks Aspen Center for Physics supported by National Science Foundation grant PHY-1607611, where a part of work was performed.
%the Center for Quantum Materials at the University of Toronto. 
This research was enabled in part by support provided by Sharcnet (\href{http://www.sharcnet.ca}{www.sharcnet.ca}) and Compute Canada (\href{http://www.computecanada.ca}{www.computecanada.ca}).
Computations were performed on the GPC and Niagara supercomputers at the SciNet HPC Consortium. 
SciNet is funded by: the Canada Foundation for Innovation under the auspices of Compute Canada; the Government of Ontario; Ontario Research Fund - Research Excellence; and the University of Toronto.

\bibliography{references}
\end{document}